\documentclass[13.5pt]{extarticle}
\usepackage[utf8]{inputenc}
\usepackage[T1]{fontenc}
\usepackage{graphicx}
\usepackage{grffile}
\usepackage{longtable}
\usepackage{wrapfig}
\usepackage{rotating}
\usepackage[normalem]{ulem}
\usepackage{amsmath}
\usepackage{textcomp}
\usepackage{amssymb}
\usepackage{capt-of}
\usepackage{hyperref}
\usepackage[margin=0.8in]{geometry}
\date{\today}
\author{Davide Magrin, Martina Capuzzo, \\ Andrea Zanella, Michele Zorzi}
\title{Proof of convergence of LoRaWAN model}
\begin{document}

\maketitle

In this document, we prove the convergence of the model proposed
in~\cite{lorawanmodel}, which aims at estimating the LoRaWAN network performance
in a single-gateway scenario. In Sec.~\ref{sec:proof}, we provide an analytical
proof of the existence of a fixed point solution for such a system. Then, in
Sec.~\ref{sec:exp}, we report experimental results, showing that the system of
the two inter-dependent equations provided by the model can be solved through
fixed-point iterations, and that a limited number of iterations is enough to
reach convergence.

\section{Proof of existence of fixed point solution}
\label{sec:proof}

For ease of writing, let $\mathbf{x}=[x_1,\dots,x_n]$ be the vector of unknowns to
be determined by solving the set of fixed point equations $f_i(\mathbf{x})=x_i$.
From the derivation of the model presented in the paper, it is apparent that
the functions $f_i()$ result from the combination of a number of continuous
(and differentiable) functions, and are themselves continuous (and
differentiable) in all the unknowns $x_i$ in the compact interval $(0,1)$.

\noindent We can apply the following reasoning iteratively.

\vspace{0.8cm}
\noindent Let’s start from $i=1$. By fixing all the parameters other than $x_1$ we can
define the functions $F_1(x_1;p_1)=f_1(x_1,…, x_i, …x_n)$ where
$p_1=[x_2,\dots,x_n]$ is the vector of parameters $\mathbf{x}$ without the
first term.

\noindent Now, analyzing the functions that yield to the expression of $F_1(x_1;
p_1)$, we can see that $F_1(x_1;p_1)$ is greater than zero when $x_1$
tends to zero, and lower than one when $x_1$ tends to one. Therefore, there must
exist a point $x_1^* \in (0,1)$ such that $F_1(x_1^*;p_1)= x_1^*$. Clearly, this
point in general depends on the parameter vector $p_1$. We hence denote by
$F_1^*(p_1)$ the fixed point solution $x_1^*$ of $F_1(x_1;p_1)$ for a certain
$p_1$. We will later prove that this function is continuous in $p_1$.

\vspace{0.8cm}
\noindent We can now define the function
$$
f_2^\circ(x_2;x _3,…,x_n)=f_2(F_1^*(x_2,x_3, \dots, x_n),x_2, x _3, \dots ,x_n).
$$

\noindent Since $f_2$ is continuous in all the parameters, and $ F_1^*$ is continuous in
$[x_2, x _3,\dots,x_n]$, then $f_2^\circ$ is also continuous in
$[x_2, x _3,\dots,x_n]$.

\noindent Furthermore, the function turns out to be greater than zero when $x_2$
tends to zero and lower than one when $x_2$ tends to one. Therefore, we can
repeat the reasoning iteratively, until we reach the function $f_n^\circ(x_n)$
that hence admits a fixed point $x_n^*$.

\noindent Hence, we get that the fixed point solution of the original problem is
given by the values $\{F_i^*(x^*_{i+1},\dots, x^*_n)\}$, for $i=1, 2, \ldots, n$.

\section*{Proof of continuity of $F_1^{*}$}
\subsection*{Theorem}

The function $F_i^*(x_{i+1},\dots, x_n)$ is continuous in $p_i^*$, where $p_i^*
= (x_{i+1}, \dots, x_n)$.

\subsection*{Proof}

\noindent We need to prove that $\forall p_i^*$, $\forall \epsilon > 0$,
$\exists \delta > 0$, such that
$$
|F_i^*(p_i^*) - F_i^*(\widetilde{p_i})| <
\epsilon, \forall \widetilde{p_i}: |p_i^* - \widetilde{p_i}| < \delta.
$$

\vspace{0.8cm}
\noindent Recalling that $F_i^*(p_i^*) = x_i^*$ such that $f^\circ(x_i^*, p_i^*)
= x_i^*$, then we can write
$$
\Rightarrow |F_i^*(p_i^*) - F_i^*(\widetilde{p_i})| = |x_i^* -
\widetilde{x_i}|
$$
where $\widetilde{x_i}$ is the fixed point of $f_i(x_i, \widetilde{p_i})$, i.e.,
$f^\circ(\widetilde{x_i}, \widetilde{p_i}) = \widetilde{x_i}$.

\vspace{0.8cm}
\noindent We hence need to prove that $\widetilde{x_i} \in (x_i^* - \epsilon,
x_i^* + \epsilon)$, i.e., that
$$
g_{\widetilde{p_i}}(x_i) = f^\circ_i (x_i, \widetilde{p_i}) - x_i = 0
$$
for some point $\widetilde{x_i} \in (x_i^* - \epsilon, x_i^* + \epsilon)
\triangleq B(x_i^*, \epsilon)$.

\vspace{0.8cm}
\noindent
Let's call $B(p_i^*, \delta)$ the ball of radius $\delta$ centered in $p_i^*$,
i.e, $B(p_i^*, \delta) \triangleq (p_i^* - \delta, p_i^* + \delta)$.

\noindent Assume, by contradiction, that $\forall \delta > 0$, $\exists \hat{p_i} \in
B(p_i^*, \delta)$ such that $g_{\widetilde{p_i}}(x_i) > 0$, $\forall x_i \in
B(x_i^*, \epsilon)$. (The case $g_{\widetilde{p_i}}(x_i) < 0$ is similar).

\noindent Let $$g_{min} = \min\limits_{x_i \in B(x_i^*, \epsilon)} g_{\widetilde{p_i}}(x_i).$$

\vspace{0.8cm}
\noindent Therefore, $g_{\widetilde{p_i}}(x_i) \geq g_{min} \, \forall x_i \in B(x_i^*,
\epsilon)$.

\vspace{0.8cm}
\noindent In particular,
$$g_{\widetilde{p_i}}(x_i^*) \geq g_{min} \qquad \Rightarrow \qquad
g_{min} \leq f^\circ_i(x_i^*, \widetilde{p_i}) - x_i^* =
f^\circ_i(x_i^*, \widetilde{p_i}) - f^\circ_i(x_i^*, p_i^*).
$$

\vspace{0.8cm}
\noindent Recalling that $f^\circ(x_i, p_i)$ is continuous, then by taking
$\hat{\epsilon} < g_{min}$, $\exists \hat{\delta}$ such that

$$|f^\circ(x_i^*, \widetilde{p_i}) - x_i^*| < \hat{\epsilon}, \forall
\widetilde{p_i} \in B(p_i^*, \hat{\delta}).$$

$$
\Rightarrow \exists \hat{\delta} > 0\textrm{, such that } \forall p \in B(p_i^*,
\hat{\delta}), \; g_{\widetilde{p_i}}(x_i^*) < g_{min} .$$
which contradicts the assumption that $g_{min}$ does not admit zeros in the ball
of radius $\epsilon$ around $x_i^*$. This concludes the proof.

\smallskip

Q.E.D.

\section{Experimental results showing system's convergence}
\label{sec:exp}

In order to provide some estimates on the convergence speed of the proposed
model, we ran a series of experiments in which we solved the model through a
fixed-point iteration procedure. First, we sampled the parameter space, varying
each parameter (such as $\alpha$, $m$, $p$, $C$, $\lambda$), thus obtaining a
set of models to solve. Then, we solved each instance of the model starting from
a randomly chosen point $\mathbf{x}$ in the solution space (i.e., the
12-dimensional space of $S_i^{UL}$ and $S_i^{DL}$, with $i \in \mathcal{SF}$).
We define
$$
\mathbf{x}_i = \left[ S_i^{UL}, S_i^{DL} \right] = \left[ S_7^{UL}, \dots,
  S_{12}^{UL}, S_7^{DL}, \dots, S_{12}^{DL} \right].
$$
We stop the fixed-point iteration when $\left\vert\left\vert \mathbf{x}_{i} -
    \mathbf{x}_{i+1}\right\vert\right\vert_{2} < 10^{-3}$, where
$\mathbf{x}_{i}$ is the solution found at the $i$-th step in the procedure, and
$\left\vert\left\vert \cdot \right\vert\right\vert_{2}$ is the Euclidean norm.

\begin{figure}
  \centering
  \includegraphics[width=0.5\linewidth]{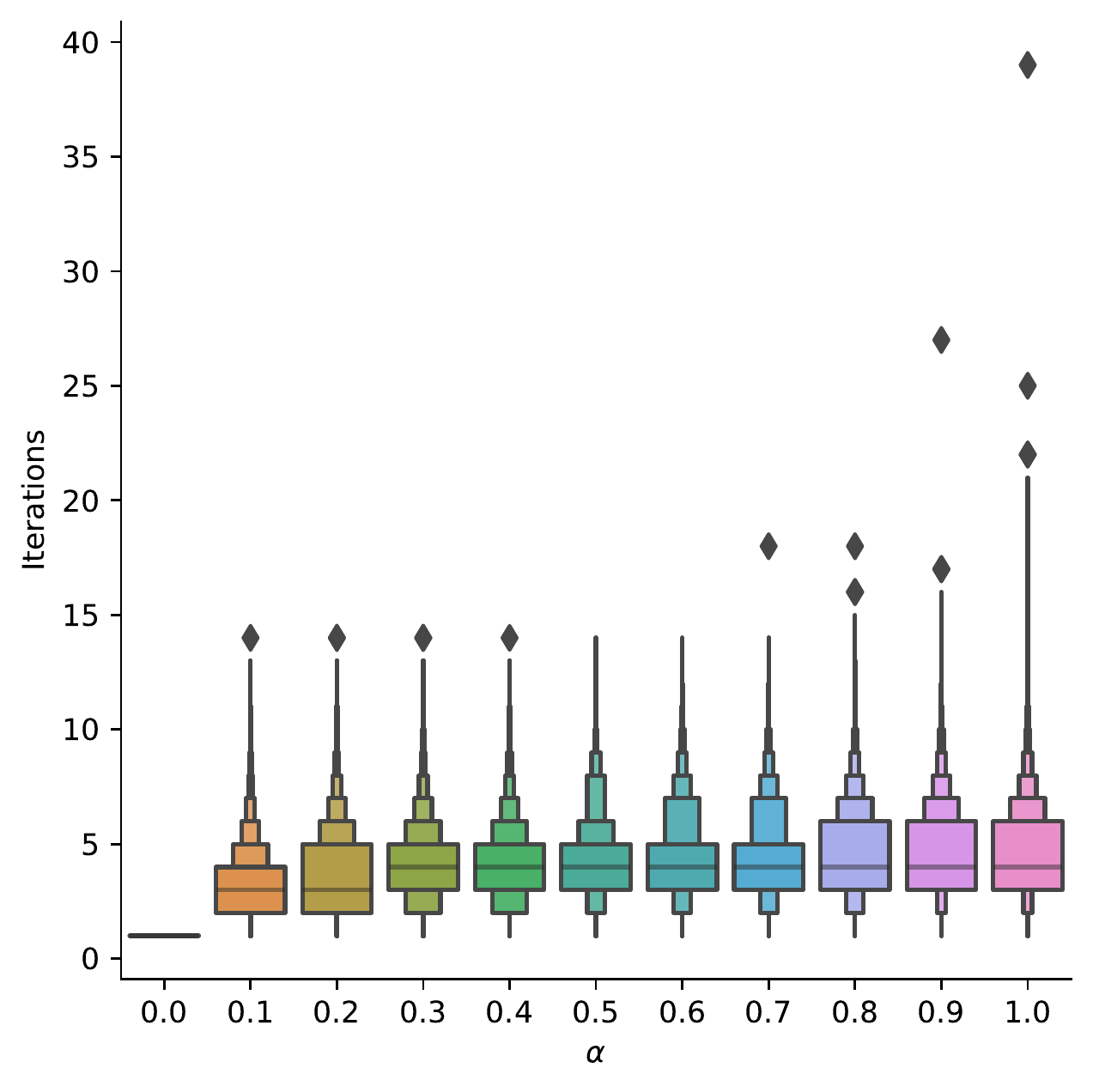}
  \caption{Distribution of the number of iterations necessary to reach
    convergence for a range of values of $\alpha$.}
  \label{fig:iterations}
\end{figure}

Figure~\ref{fig:iterations} contains the results of such analysis: the box plots
show the distribution of the number of iterations necessary to reach
convergence. The data is plotted here for various values of $\alpha$, to
highlight how the number of iterations might depend on the value of some
parameters, while never exceeding 40 in the worst case, ensuring, thus, a quick
convergence for all the explored combinations of the parameters and choice of
the initialization point.

\end{document}